\documentclass[submission,copyright,creativecommons]{eptcs}
\providecommand{\event}{4th Workshop on Formal Integrated Development Environment (F-IDE 2018)} 
\usepackage{breakurl}             
\usepackage{underscore}           

\usepackage{verbatim}
\usepackage{xspace}
\usepackage{graphicx}
\usepackage{booktabs}
\usepackage{paralist}
\usepackage{rotating}

\usepackage{hyperref}
\hypersetup{
colorlinks = true, 
linkcolor=black, 
citecolor=black, 
urlcolor=black 
}

\usepackage{xspace}

\newcommand{\asmeta}{ASMETA\xspace}
\newcommand{\asmetaF}{{\tt AsmetaF}\xspace}

\newcommand{\mcr}{{\tt MCR}\xspace}
\newcommand{\fr}{{\tt FR}\xspace}
\newcommand{\chr}{{\tt ChR}\xspace}
\newcommand{\ar}{{\tt AR}\xspace}
\newcommand{\lr}{{\tt LR}\xspace}
\newcommand{\car}{{\tt CaR}\xspace}
\newcommand{\nr}{{\tt NR}\xspace}
\newcommand{\tsimpl}{{\tt TS}\xspace}
\newcommand{\rs}{{\tt RS}\xspace}

\usepackage[]{listings}
\lstdefinelanguage{AsmetaL} 
{morekeywords={par, endpar, if, endif, then, else, seq, endseq, signature, definitions, asm, import, function, domain, main, rule, macro, invariant, over, choose, otherwise, forall, exists, unique, let, endlet, with, ifnone, abstract, default, init, do, agent, dynamic, controlled, monitored, in, out, static, derived, subsetof, switch, case, endswitch, enum, CTLSPEC, LTLSPEC},
sensitive=true, morecomment=[l]{//}, morecomment=[s]{/*}{*/},
morecomment=[l][\color{white}\tiny]{'},
morestring=[b]",tabsize=2, columns=fullflexible, basicstyle=\footnotesize\sffamily, captionpos=b}

\newcounter{researchquestionCount}
\newcommand{\researchquestion}[1]{\stepcounter{researchquestionCount}\begin{itemize}\item [\textbf{RQ\arabic{researchquestionCount}:}] \emph{#1}\end{itemize}\vspace{-5pt}}

\title{AsmetaF: A Flattener for the \asmeta Framework\thanks{The first two authors are supported by ERATO HASUO Metamathematics for Systems Design Project (No. JPMJER1603), JST. Funding Reference number: 10.13039/501100009024 ERATO.}}
\author{Paolo Arcaini
\institute{National Institute of Informatics\\Japan}
\email{arcaini@nii.ac.jp}
\and Riccardo Melioli \qquad\qquad Elvinia Riccobene
\institute{Dipartimento di Informatica, Universit\`{a} degli Studi di Milano\\Italy}
\email{riccardo.melioli@studenti.unimi.it \quad elvinia.riccobene@unimi.it}
}
\def\titlerunning{AsmetaF: A Flattener for the \asmeta Framework}
\def\authorrunning{P. Arcaini, R. Melioli \& E. Riccobene}
\begin{document}
\maketitle

\begin{abstract}
Abstract State Machines (ASMs) have shown to be a suitable high-level specification method for complex, even industrial, systems; the \asmeta framework, supporting several validation and verification activities on ASM models, is an example of a formal integrated development environment. Although ASMs allow modeling complex systems in a rather concise way --and this is advantageous for specification purposes--, such concise notation is in general a problem for verification activities as model checking and theorem proving that rely on tools accepting simpler notations.

In this paper, we propose a \emph{flattener} tool integrated in the \asmeta framework that transforms a general ASM model in a \emph{flattened} model constituted only of update, parallel, and conditional rules; such model is easier to map to notations of verification tools. Experiments show the effect of applying the tool to some representative case studies of the \asmeta repository.
\end{abstract}

\section{Introduction}
Abstract State Machines~\cite{ASMbook,bookModelingCompanionASM2018} (ASMs) is a formal specification method based on model refinement, which has been used in several application domains and case studies~\cite{LGSjournalSTTT2017,abz2016siSCICO2017,AsmCloudFAOC2016}. To overcome the lack of tool support and foster the use of ASMs for rigorous software development, in 2008 we started the \asmeta (ASM mETAmodeling) project with the goal of developing a set of tools supporting different activities of the ASM-based system development process~\cite{LGSjournalSTTT2017}, and operating in an integrated way to reuse model information. Today, \asmeta~\cite{modelDrivenProcess} exists as a framework for specification, validation (by simulation, scenario construction, model-based testing) and verification (static analysis, model checking, symbolic verification, refinement correctness, runtime verification), as well as automatic code generation. Exploiting the Model-Driven methodology (which is at the base of the whole framework development), some of these tools have been developed from scratch, while many others have been obtained by mapping ASMs (usually using \emph{model2model} or \emph{model2text} transformations) into the native formalisms of already existing tools (e.g., model checkers and SMT solvers) in order to exploit their functionalities. Our whole project is, indeed, based on the idea that the \asmeta tool-set should be a formal \emph{integrated} development environment for ASMs. However, the integration of tools into the \asmeta framework has caused some difficulties that justify the work we present here and that we motive in the following.

\noindent {\bf Motivation}
ASMs use a plain mathematical notation to model a system configuration (i.e., a \emph{state}) in terms of a mathematical algebra, and use a set of powerful rule constructors to specify system behavior (i.e., state \emph{transitions}). ASMs provide, therefore, a powerful language that permits to describe complex systems in a rather concise way. Although it is an advantage when modeling, this notational conciseness can be a problem for tools integration: target languages have their own syntax and semantics, and translating an ASM to a target model by maintaining the intended computational model is not a trivial work; moreover, ASM specifications must often be translated to less expressive languages, and implementing these transformation tools is rather complicated, as the semantics of the complex rule constructors of the ASM language must be taken into account and guaranteed.
 
In the past, different mappings have been developed to model checkers as SPIN~\cite{spinASM} and NuSMV~\cite{asmetaSMV}, to SMT solvers~\cite{smtAFM2017,smtBasedAsmRefProverSEFM2016}, and to C++ code~\cite{asmC}. All these target notations, although can in principle represent the same class of systems as ASMs, have syntaxes that are very different from the ASM notation, with less expressive constructs; therefore, the integration of these tools into \asmeta usually supports only specific classes of ASMs. Some constructs of the ASM formalism are indeed difficult to translate in the target notation, and, although possible, we did not implement such translations because too much complex (e.g., the mappings to model checkers NuSMV and SPIN do not support variable arguments in functions). On the other hand, we observe that tools integrated into \asmeta usually perform similar pre-processing of supported ASM constructs (e.g., translation of unbounded parallelism of the forall rule is usually implemented by an unfolding), and that this pre-processing could be extracted from the integrated tools and made separately available for all the integrations.

This necessity was again confirmed by our recent work on devising a new mapping to the probabilistic model checker PRISM\footnote{\url{http://www.prismmodelchecker.org/}} that will be used for ASM-based analysis of cyber-physical systems in the context of the ERATO MMSD project~\cite{Hasuo2017}: we realized that instead of trying to directly map any ASM in PRISM (that provides an extremely limited language), it would have been better to go through a {\it simpler}, but still equivalent, ASM that uses a limited set of ASM constructs. Such simpler ASM would have been as the result of the pre-processing phase of other integrated tools.

\noindent{\bf Contribution}
To simplify the porting of ASMs towards other modeling languages, to reuse tools for model validation and verification, and to foster tool integration into the \asmeta framework, we here propose a \emph{flattener} (\asmetaF) that, given an ASM model $M$, produces an equivalent model $M_f$ that only contains update, conditional, and parallel rules; we consider the $M_f$ model to be in a \emph{normal form}. The idea is that translating the normal form to the target languages of verification frameworks (e.g., NuSMV, SMT-LIB) or code is much easier than considering ASMs containing any possible construct. We are currently using \asmetaF in the development of the mapping of ASM to PRISM. Moreover, we have integrated the tool with the AsmetaSMV tool; this has allowed us to support a wider set of specifications, namely those having variable function arguments. In the future, we will integrate it in existing verification tools of the \asmeta framework and new tools requiring flattening.

\noindent {\bf Paper structure} Sect.~\ref{sec:asms} presents some background on the ASM method. Sect.~\ref{sec:flattener} introduces the flattener transformations, and describes how we validated the approach. Sect.~\ref{sec:experiments} describes some preliminary experiments, Sect.~\ref{sec:related} reviews some related work, and Sect.~\ref{sec:conclusions} concludes the paper.

\section{Abstract State Machines}\label{sec:asms}
\emph{Abstract State Machines} (ASMs)~\cite{ASMbook,bookModelingCompanionASM2018} are transition systems based on the concept of \emph{state} representing the instantaneous system configuration, and \emph{transition rules} describing the change of state.

ASM \emph{states} are multi-sorted first-order structures, i.e., domains of objects with functions and predicates defined on them. An ASM state $S$ is represented by a set of couples (\emph{location}, \emph{value}). ASM \emph{locations}, namely pairs (\emph{function-name}, \emph{list-of-parameter-values}), represent the abstract ASM concept of basic object containers (memory units). Location \emph{updates} represent the basic units of state change and they are given as assignments, each of the form $\mathit{loc} := v$, where $loc$ is a location and $v$ its new value.

ASM \emph{transition rules} express how function interpretations are modified from one state to the next one, and therefore describe the system configuration changes. The basic form of a transition rule is the {\tt conditional rule}: ``\textbf{if} \emph{Condition} \textbf{then} \emph{Updates}'', where \emph{Updates} is a set of function updates (or {\tt update rules}) of the form $f(t_{1}, \ldots, t_{n}):= t$ which are simultaneously executed when \emph{Condition} is true; $f$ is an $n$-ary function and $t_{1}, \ldots, t_{n}, t$ are terms. Due to their parallel execution, we require updates to be consistent, i.e., no pair of updates can simultaneously update the same location to different values.

Besides update and conditional, there is a finite set of \emph{rule constructors} to model submachine calls (\texttt{macro (call)} rule), simultaneous parallel actions (\texttt{par} rule), non-determinism (\texttt{choose} rule), unrestricted synchronous parallelism (\texttt{forall} rule), abbreviation on terms of rules (\texttt{let} rule). There are also derived rule constructors, as the \texttt{case} rule that is defined as alternative disjointed guarded rules.\footnote{All above mentioned rule constructors are characteristics of the \emph{basic ASMs}, which dispose of potentially unrestricted non-determinism and parallelism (appearing in the form of the \emph{choose} and \emph{forall} rules) and to distinguish a version with flat specifications from structured versions (by using the \emph{macro call} rule). Besides basic ASMs, there are advanced classes of ASMs having mechanisms for domain extention (\texttt{extend} rule), action sequentialization (\texttt{seq} rule), and invocation of sub-machines reporting values. In the current work, we do not consider such advanced ASM classes that will be addressed as future work.}

Functions remaining unchanged during the computation are \emph{static}. Those updated by agent actions are \emph{dynamic}, and distinguished in \emph{monitored} (read by the machine and modified by the environment) and \emph{controlled} (read and written by the machine).

A \emph{computation} of an ASM is a finite or infinite sequence $S_0, S_1,$ $\ldots, S_n, \ldots$ of machine states, where $S_0$ is an initial state and each $S_{n + 1}$ is obtained from $S_n$ by simultaneously firing all the transition rules which are enabled in $S_n$. The (unique) \emph{main rule} is a transition rule and represents the starting point of the computation. An ASM can have more than one \emph{initial state}. It is possible to specify state \emph{invariants}.

A \emph{multi-agent ASM} models concurrent and distributed computations. It is defined as a set of pairs $ M = \{(a, \emph{ASM}(a)) \} $ where $a$ is an element of a predefined set \emph{Agent}, and $\emph{ASM}(a)$ is a machine specifying its behavior. A predefined function \emph{program} on \emph{Agent} associates an agent with its ASM, and a special function $\mathit{self}:\mathit{Agent}$, interpreted by each agent $a$ as itself, allows for self-reference in transition rules.

\asmeta\footnote{\url{http://asmeta.sourceforge.net/}}~\cite{modelDrivenProcess} is a tool-set for ASMs, which provides basic functionalities for specification and model analysis techniques (validation, verification, testing, model review, requirements analysis, runtime monitoring, etc.). AsmetaL is the textual notation to encode ASM models into \asmeta.

\section{Flattener}\label{sec:flattener}

In order to improve tools integration in \asmeta and to overcome some shortcomings due to the high level and concise mathematical notation of the ASMs w.r.t. less expressive (in terms of conciseness) formalisms of the integrated tools, we developed a \emph{flattener}. Given an ASM $M$ written in {\it general form}---i.e., containing any kind of rule and any level of nesting---, the flattener produces an ASM $M_f$ in \emph{normal form} (if all the flattener transformation rules are applied). An ASM is in normal form if, in the main rule, it only contains a parallel rule composed of a set of update rules and conditional rules (without else branch); each conditional rule must contain either an update rule or a parallel of update rules.

The flattener applies a series of transformations shown in Table~\ref{table:flattenerTransormations} and described in the following.
\begin{table}[!tb]
\center
\begin{tabular}{p{0.02\textwidth}|p{.36\textwidth}|p{.54\textwidth}}
\toprule
& \textbf{Original ASM} & \textbf{Flattened ASM}\\
\midrule
\begin{turn}{90}\mcr\end{turn} &
\vspace{-22pt}
\begin{minipage}[t]{\textwidth}
\begin{lstlisting}[mathescape, language=AsmetaL]
rule r(v$_1$ in D$_1$, $\ldots$,v$_n$ in D$_n$) = R[v$_1$,...,v$_n$]
$\ldots$
r[t$_1$,...,t$_n$] //macro call rule
\end{lstlisting}
\end{minipage}
&
\vspace{-22pt}
\begin{minipage}[t]{\textwidth}
\begin{lstlisting}[mathescape, language=AsmetaL]
//Macro rule r is removed

R[v$_1$ $\mapsto$ t$_1$, $\ldots$, v$_n$ $\mapsto$ t$_n$]
\end{lstlisting}
\end{minipage}\\
\hline
\begin{turn}{90}\mcr\end{turn} &
\vspace{-22pt}
\begin{minipage}[t]{\textwidth}
\begin{lstlisting}[mathescape, language=AsmetaL]
rule rAgentKind = R[self]
$\ldots$
program(a) //a is an AgentKind agent
$\ldots$
agent AgentKind: rAgentKind[]
\end{lstlisting}
\end{minipage}
&
\vspace{-22pt}
\begin{minipage}[t]{\textwidth}
\begin{lstlisting}[mathescape, language=AsmetaL]
//Macro rule rAgentKind is removed

R[self $\mapsto$ a]

//Program declaration for AgentKind is removed
\end{lstlisting}
\end{minipage}\\
\hline
\begin{turn}{90}\fr\end{turn}
&
\vspace{-5pt}
\begin{minipage}[t]{\textwidth}
\begin{lstlisting}[mathescape, language=AsmetaL]
forall v$_1$ in D$_1$, $\ldots$, v$_n$ in D$_n$
with guard[v$_1$, $\ldots$, v$_n$] do
	R[v$_1$, $\ldots$, v$_n$]
\end{lstlisting}
\end{minipage}
&
\vspace{-5pt}
\begin{minipage}[t]{\textwidth}
{\footnotesize\sffamily (d$_1^1$,$\ldots$,d$_n^1$), \ldots, (d$_1^m$, $\ldots$, d$_n^m$) $\in$ 	D$_1$$\times$$\ldots$$\times$D$_n$
	with $m = \prod_{j=1}^n |D_j|$}
\begin{lstlisting}[mathescape, language=AsmetaL]
par
	if guard[v$_1$ $\mapsto$ d$_1^1$, $\ldots$, v$_n$ $\mapsto$ d$_n^1$] then
		R[v$_1$ $\mapsto$ d$_1^1$, $\ldots$, v$_n$ $\mapsto$ d$_n^1$]
	endif
	...
endpar
\end{lstlisting}
\end{minipage}\\
\hline
\begin{turn}{90}\chr\end{turn} &
\vspace{-20pt}
\begin{minipage}[t]{\textwidth}
\begin{lstlisting}[mathescape, language=AsmetaL]


choose v in D with guard[v] do
	R[v]
[ifnone R$_{\mathit{none}}$]
\end{lstlisting}
\end{minipage}
&
\vspace{-20pt}
\begin{minipage}[t]{\textwidth}
\begin{lstlisting}[mathescape, language=AsmetaL]
function f$_{\mathit{choose}}$ = chooseone({v in D | guard[v] : v})

if isDef(f$_{\mathit{choose}}$) then R[v $\mapsto$ f$_{\mathit{choose}}$]
[else R$_{\mathit{none}}$]
endif
\end{lstlisting}
\end{minipage}\\
\hline
\begin{turn}{90}\ar\end{turn} &
\vspace{-15pt}
\begin{minipage}[t]{\textwidth}
\begin{lstlisting}[mathescape, language=AsmetaL]
f($t_1$, $\ldots$, $t_n$)
\end{lstlisting}
\end{minipage}
&
\vspace{-15pt}
\begin{minipage}[t]{\textwidth}
\begin{lstlisting}[mathescape, language=AsmetaL]
let (v$_1$ = t$_1$, $\ldots$, v$_n$ = t$_n$) in f(v$_1$, $\ldots$, v$_n$) endlet
\end{lstlisting}
\end{minipage}\\
\hline
\begin{turn}{90}\lr\end{turn}&
\vspace{-5pt}
\begin{minipage}[t]{\textwidth}
\begin{lstlisting}[mathescape, language=AsmetaL]
let (v$_1$=t$_1$, $\ldots$, v$_n$=t$_n$) in
	R[v$_1$, $\ldots$, v$_n$]
endlet
\end{lstlisting}
\end{minipage}
&
\vspace{-5pt}
\begin{minipage}[t]{\textwidth}
{\footnotesize\sffamily D$_1$, $\ldots$, D$_n$ are the domains of $t_1, \ldots, t_n$ and\\
(d$_1^1$,$\ldots$,d$_n^1$), \ldots, (d$_1^m$, $\ldots$, d$_n^m$) $\in$ D$_1$$\times$$\ldots$$\times$D$_n$
	with $m = \prod_{j=1}^n |D_j|$}
\begin{lstlisting}[mathescape, language=AsmetaL]
par
	if t$_1$ = d$_1^1$ and $\dots$ and t$_n$ = d$_n^1$ then R[v$_1$ $\mapsto$ d$_1^1$, $\ldots$, v$_n$ $\mapsto$ d$_n^1$]
	endif
	$\dots$
endpar
\end{lstlisting}
\end{minipage}\\
\hline
\begin{turn}{90}\car\end{turn} &
\vspace{-20pt}
\begin{minipage}[t]{\textwidth}
\begin{lstlisting}[mathescape, language=AsmetaL]
switch(t)
	case t$_1$: R$_1$
	$\ldots$
	case t$_n$: R$_n$
	[otherwise R$_{o}$]
endswitch
\end{lstlisting}
\vspace{-10pt}
\end{minipage}
&
\vspace{-20pt}
\begin{minipage}[t]{\textwidth}
\begin{lstlisting}[mathescape, language=AsmetaL]
par
	if t = t$_1$ then R$_1$ endif
	$\ldots$
	if t = t$_n$ then R$_n$ endif
	[if t != t$_1$ and $\ldots$ and t != t$_n$ then R$_{o}$ endif]
endpar
\end{lstlisting}
\end{minipage}\\
\hline
\begin{turn}{90}\nr\end{turn} &
\vspace{-15pt}
\begin{minipage}[t]{\textwidth}
\begin{lstlisting}[mathescape, language=AsmetaL]
if guard$_1$ then
	if guard$_2$ then R$_t$
	else R$_e$ endif
endif
\end{lstlisting}
\end{minipage}
&
\vspace{-15pt}
\begin{minipage}[t]{\textwidth}
\begin{lstlisting}[mathescape, language=AsmetaL]
par
	if guard$_1$ and guard$_2$ then R$_t$ endif
	if guard$_1$ and not(guard$_2$) then R$_e$ endif
endpar
\end{lstlisting}
\end{minipage}\\
\bottomrule
\end{tabular}
\caption{Flattener transformations}
\label{table:flattenerTransormations}
\end{table}

\vspace{5pt}

\noindent {\bf \mcr: Macro Call rule Remover} A {\it macro rule} is a named rule $r$ with some formal parameters $v_1$, \ldots, $v_n$, and a rule body $R$ defined in terms of the parameters. A {\it macro call rule} is an invocation of rule $r$ with actual parameters $t_1$, \ldots, $t_n$. The flattener transformation \mcr replaces each occurrence of a call rule $r$ with the macro rule body $R$; occurrences of formal parameters in the rule are replaced by the actual parameters used in the macro call rule. In multi-agent ASMs, given a specific subset ${\it AgentKind}$ of \emph{Agent}, a macro rule ${\it rAgentKind}$ specifies the program of all agents in ${\it AgentKind}$, and, by the keyword ${\it program}$, it is possible to invoke the program of an agent $a$ in ${\it AgentKind}$; in rule ${\it rAgentKind}$, the keyword ${\it self}$ is used to reference the current agent executing the rule. \mcr flattens also program invocations; an invocation ${\it program(a)}$ is replaced with the rule $R$ (body of the agent rule), where each occurrence of ${\it self}$ is replaced with $a$. At the end, all the macro rules declared in the ASM model are removed.

\vspace{5pt}

\noindent {\bf \fr: Forall rule Remover} In a \emph{forall rule}, the rule $R$ is executed in parallel with all the values of variables $v_1$, \ldots, $v_n$ satisfying the \emph{guard}. The flattener transformation \fr, for each combination $\overline{d} = (d_1, \ldots, d_n)$ of values of the domains $D_1$, \ldots, $D_n$, 
builds a conditional rule (without else branch)\footnote{Note that in AsmetaL the domains of a forall must be finite, so the number of generated conditional rules will be finite.} whose guard is that of the forall rule, instantiated over values $\overline{d}$ (i.e., variables $v_1$, \ldots, $v_n$ are replaced by values $d_1$, \ldots, $d_n$); in a similar way, the rule in the {\it then} branch is the rule $R$ of the forall body instantiated over $\overline{d}$.

\vspace{5pt}

\noindent{\bf \chr: Choose rule Remover}
In a \emph{choose rule}, the rule $R$ is executed with a value of $v$, nondeterministically chosen, that satisfies \emph{guard}. If such value does not exist, the choose rule does nothing. The flattener transformation \chr embeds the non-deterministic choice in a derived function $f_{\it choose}$ that randomly selects one of the values of the choose domain; the rule is replaced by a conditional rule that checks whether $f_{\it choose}$ is defined (i.e., it is possible to select a value from the domain) and, if so, calls $R$ instantiated over $f_{\it choose}$. In a choose rule, it is also possible to specify a rule $R_{\it none}$ that must be executed when no choice can be performed; in the flattened version, this rule is reported in the {\it else} branch.

\vspace{5pt}

\noindent {\bf \ar: Arguments Remover}
Function locations are identified at runtime by interpreting the terms used as function arguments. Such feature is usually particularly difficult to handle in target notations; NuSMV, for example, allows to specify arrays (that could be used to model functions), but does not allow to dynamically accessing them. The flattener transformation \ar removes terms used as function arguments and replaces them by suitable let rules (that can then be flattened by the flattener transformation \lr).

\vspace{5pt}

\noindent {\bf \lr: Let rule Remover}
A {\it let rule} associates logical variables $v_1$, \ldots, $v_n$ to terms $t_1$, \ldots, $t_n$; the rule body $R$ is defined in terms of the variables. The flattener transformation \lr removes the rule by considering all the possible values assumed by the terms; for each combination $\overline{d} = (d_1, \ldots, d_n)$ of values of the terms domains, a conditional rule is created: the guard checks whether the terms assume the values in $\overline{d}$, and the {\it then} rule is the rule body $R$ of the let rule, instantiated over $\overline{d}$.

\vspace{5pt}

\noindent {\bf \car: Case rule Remover}
In a case rule, a term $t$ is compared with terms $t_1$, \ldots, $t_n$, each one associated with a rule $R_i$ to be executed if $t$ evaluates as $t_i$. An optional {\it otherwise} branch can specify a rule $R_o$ to execute when $t$ does not match to any of the $t_i$. The flattener transformation \car introduces a parallel of conditional rules, each checking whether $t$ is equal to $t_i$ and then executing the corresponding rule $R_i$ in the then branch. An additional conditional rule is added if the {\it otherwise} branch is present.

\vspace{5pt}

\noindent {\bf \nr: Nesting Remover}
A nested conditional rule is replaced by parallel conditional rules, by unfolding the rules and aptly combining their guards.

\vspace{5pt}

\noindent {\bf Simplifier}
Applying the previous flattener transformations could produce some terms only containing constants; such terms can be evaluated statically at parsing time. Therefore, in order to avoid unnecessary rules in the flattened models, after the application of a flattener transformation, we apply two {\it simplifiers}:
\begin{compactenum}
\item \tsimpl visits all the terms and, if possible, evaluates them or simplifies them; for example, a function term $\mathit{and}(a, \mathit{true})$ is simplified to $a$, $3 < 4$ is simplified to $\mathit{true}$, $2 + 1$ is simplified to 3, and so on; \tsimpl can simplify logical, mathematical, and relational terms;
\item \rs visits all the rules and, if possible, removes or simplifies them; for example a conditional rule with guard equal to $\mathit{true}$ is replaced with its {\it then} rule.
\end{compactenum}

\paragraph{Application order of flattener transformations}
All the flattener transformations are applied in the order in which they have been presented. The order guarantees that no construct that should be flattened is not. Indeed, a transformation could introduce some constructs that are further flattened by another one; namely, \lr must be executed after \ar because \ar introduces let rules that are then flattened by \lr; in a similar way, \nr must be applied after all the other transformations because it must remove the nesting they introduce. However, the chosen order is not the only possible; indeed, although there are couples of transformations that must be executed in a given order, the order of other couples could be exchanged. We will discuss about the {\it best} order in the experiments (see {\bf RQ1} in Sect.~\ref{sec:experiments}).

\paragraph{Tool implementation}
The flattener has been implemented in the tool \asmetaF. The tool has been designed in a modularized way such that the user can decide which flattener transformations to apply; in some cases, it may be not necessary to flatten all the ASM constructs, as some of them could be natively supported by the target language. For example, a programming language as C supports nesting, and so it is not necessary to remove it. The tool is meant to be used as pre-processing step of other tools. However, we provide a standalone version for demonstration purposes.\footnote{A jar file of the tool can be downloaded from \url{https://goo.gl/vShfbJ}}

\subsection{Validation of the approach}\label{sec:validation}

The proposed flattener transformations preserve the ASM semantics; however, it could be that their implementation in \asmetaF is not correct. In order to guarantee the correctness of \asmetaF, we should prove semantic equivalence between original and flattened models, but this is in general difficult to achieve. Therefore, we perform two kinds of validation, {\it syntactic} and {\it semantic}.

In \emph{syntactic validation}, we simply check whether the produced flattened ASM is syntactically correct, i.e., it can be parsed correctly by the \asmeta parser.

In the \emph{semantic validation}, by means of model checking and scenario-based validation, we try to check that the semantics of the model is preserved. We use the AsmetaSMV tool to check that the temporal properties specified in the original model are equally evaluated in the flatten model. The tool AsmetaV, instead, allows to write {\it scenarios} (similar to test cases) that drive the model simulation and check that the ASM state (values of controlled locations) is as expected; we run the scenarios written for the original model also on the flattened model and we expect that it passes them. 

In the future, we plan to devise a more systematic way to perform validation. For example, we could automatically produce scenarios achieving rule coverage of the original and target models: the target model should pass scenarios generated for the original one (to check that the flattener preserves the behavior), and the original model should pass scenarios generated for the target one (to check that the flattener does not introduce additional behaviors).

\section{Experiments}\label{sec:experiments}

We applied all the transformations to 13 representative models of the \asmeta repository\footnote{All the original and the flattened models, together with the scenarios used for validation, are available at \url{http://fmse.di.unimi.it/sw/FIDE2018AsmetaF.zip}} as a Landing Gear System~\cite{LGSjournalSTTT2017}, a hemodialysis device~\cite{abz2016siSCICO2017}, a device for measuring amblyopia, and a termination detection algorithm by Dijkstra (from Dagstuhl Seminar 13372\footnote{\url{https://www.dagstuhl.de/de/programm/kalender/semhp/?semnr=13372}}). Note that some of the case study models were obtained by refinement and we took the last refined model.
Table~\ref{table:benchSize} reports, for all the models, the number of their rules.
\begin{table}[!tb]
\centering
\resizebox{0.98\textwidth}{!}{%
\begin{tabular}{lrrrrrrrrr}
\toprule
& \multicolumn{9}{c}{Rule}\\
\cline{2-10}
Model & Update & Parallel & Conditional & Forall & Choose & Case & Let & MacroCall & All\\
\midrule
CoffeeVendingMachine & 2 & 1 & 3 & 0 & 1 & 0 & 0 & 2 & 9\\
\hline
DijkstraTermination & 9 & 6 & 8 & 1 & 3 & 0 & 0 & 9 & 36\\
\hline
ferrymanSimulator\_raff1 & 5 & 1 & 3 & 0 & 0 & 0 & 0 & 2 & 11\\
\hline
GameOfLife & 2 & 0 & 3 & 1 & 0 & 0 & 0 & 1 & 7\\
\hline
GilbreathCardTrick & 15 & 5 & 7 & 2 & 3 & 1 & 0 & 9 & 42\\
\hline
HemodialysisRef3 & 146 & 78 & 228 & 1 & 0 & 0 & 0 & 192 & 645\\
\hline
LandingGearSystem\_3L & 38 & 15 & 9 & 0 & 0 & 5 & 0 & 4 & 71\\
\hline
OneWayTrafficLight & 5 & 9 & 8 & 0 & 0 & 0 & 0 & 16 & 38\\
\hline
PetriNet & 1 & 0 & 0 & 1 & 1 & 0 & 0 & 1 & 4\\
\hline
philosophers1 & 6 & 3 & 4 & 0 & 1 & 0 & 0 & 3 & 17\\
\hline
Roulette & 4 & 2 & 3 & 0 & 1 & 0 & 0 & 4 & 14\\
\hline
SluiceGateMotorCtl & 9 & 7 & 8 & 0 & 0 & 0 & 0 & 4 & 28\\
\hline
StereoacuityRaff5 & 20 & 6 & 11 & 0 & 0 & 0 & 0 & 15 & 52\\
\hline
\hline
AVG & 20.15 & 10.23 & 22.69 & 0.46 & 0.77 & 0.46 & 0 & 20.15 & 74.92\\
\bottomrule
\end{tabular}%
}
\caption{Benchmarks size}
\label{table:benchSize}
\end{table}
The table also reports, for each kind of rule, the average number among the models. We observe that the update, the conditional, and the macro call rules are the most used ones.

\researchquestion{Which are the most applied flattener transformations?}
We are here interested in finding which are the transformations that are applied more often. Table~\ref{table:appliedFlatteners} reports how many times each transformation is applied to each model.
\begin{table}[!tb]
\centering
\resizebox{0.98\textwidth}{!}{
\begin{tabular}{lrrrrrrr|rr|r}
\toprule
& \multicolumn{7}{c}{Flattener transformation} & \multicolumn{2}{c}{Simplifier}\\
\cline{2-8} \cline{9-10} \cline{11-11}
Model & \mcr & \fr & \chr & \ar & \lr & \car & \nr & \tsimpl & \rs & Time (sec)\\
\midrule
CoffeeVendingMachine & 2 & 0 & 1 & 1 & 1 & 0 & 2 & 0 & 0 & 0.01\\
\hline
DijkstraTermination & 9 & 1 & 18 & 12 & 42 & 0 & 5 & 0 & 0 & 0.11\\
\hline
ferrymanSimulator\_raff1 & 2 & 0 & 0 & 3 & 9 & 0 & 2 & 1 & 1 & 0.08\\
\hline
GameOfLife & 1 & 1 & 0 & 0 & 0 & 0 & 2 & 0 & 0 & 0.02\\
\hline
GilbreathCardTrick & 9 & 4 & 3 & 50 & 50 & 1 & 4 & 12 & 0 & 0.37\\
\hline
HemodialysisRef3 & 192 & 1 & 0 & 0 & 0 & 0 & 8 & 0 & 0 & 0.94\\
\hline
LandingGearSystem\_3L & 4 & 0 & 0 & 0 & 0 & 6 & 4 & 0 & 0 & 0.03\\
\hline
OneWayTrafficLight & 16 & 0 & 0 & 12 & 20 & 0 & 1 & 96 & 16 & 0.04\\
\hline
PetriNet & 1 & 1 & 1 & 4 & 4 & 0 & 1 & 0 & 0 & 0.01\\
\hline
philosophers1 & 3 & 0 & 1 & 10 & 5414 & 0 & 2 & 20102 & 1802 & 59.93\\
\hline
Roulette & 4 & 0 & 1 & 1 & 1 & 0 & 3 & 37 & 37 & 1.36\\
\hline
SluiceGateMotorCtl & 4 & 0 & 0 & 4 & 4 & 0 & 1 & 2 & 2 & 0.01\\
\hline
StereoacuityRaff5 & 15 & 0 & 0 & 0 & 0 & 0 & 6 & 0 & 1 & 0.06\\
\hline
\hline
AVG & 20.15 & 0.62 & 1.92 & 7.46 & 426.54 & 0.54 & 3.15 & 2892.86 & 265.57 & 4.84\\
\bottomrule
\end{tabular}}
\caption{Applied flattener transformations and execution time}
\label{table:appliedFlatteners}
\end{table}
Since \mcr is used at the beginning, it is applied exactly the same number of times as the number of macro call rules (see Table~\ref{table:benchSize}); note that, although \mcr could be applied at any stage during the flattening process, it makes sense to use it at the beginning since it is applied so many times (it is the second most used transformation). Applying it after some other transformations (e.g., \fr) would probably increase even more the number of times it is used.

The most used transformation is \lr; although the original models do not contain any let rule, these are introduced by \ar. Note that in some models (e.g., philosophers1) the number of applications of \lr is much higher than that of \ar, because the let rules are nested: during the flattening, the inner let rule is visited as many times as the number of conditional rules created by outer let rule.

The value reported for \nr is the difference between the maximum nestings of the starting model and of the flattened one (i.e., how many nesting levels have been removed). We observe that, on average, 3.15 levels of nesting are removed, meaning that the developers tend to write quite nested models.

\researchquestion{Does the simplification have any effect?}
We here check whether the simplification of terms and rules (embedded in all the flattener transformations) has some effect. Table~\ref{table:appliedFlatteners} reports, for each model, the number of simplifications performed by the two simplifiers. We observe that, for more than half of the models, the simplifications are actually applied. For example, in the flattening of OneWayTrafficLight and philosophers1, several terms are simplified; this is due to the fact that both models contain several guards of conditional rules that depend on functions with dynamic arguments. When these arguments are flattened by \ar and \lr, some resulting guards can be simplified by \tsimpl either to {\it true} or to {\it false}; as a consequence of the simplifications of guards, some conditional rules can be simplified by \rs, either by removing them (if the guard is {\it false}) or by replacing them with the {\it then} branch (if the guard is {\it true}).

\researchquestion{Which is the effect of flattening?}
We are now interested in evaluating the effect of applying the flattener to the models. Table~\ref{table:sizeFlattened} reports the size of the flattened models in terms of number of update, parallel, and conditional rules.
\begin{table}[!tb]
\centering
\begin{tabular}{lrrrrr}
\toprule
& \multicolumn{5}{c}{Rule}\\
\cline{2-6}
Model & Update & Parallel & Conditional & All & $\Delta$ \%\\
\midrule
CoffeeVendingMachine & 6 & 2 & 5 & 13 & 44\%\\
\hline
DijkstraTermination & 276 & 13 & 252 & 541 & 1403\%\\
\hline
ferrymanSimulator\_raff1 & 145 & 1 & 145 & 291 & 2545\%\\
\hline
GameOfLife & 32 & 1 & 32 & 65 & 829\%\\
\hline
GilbreathCardTrick & 900 & 2 & 898 & 1800 & 4186\%\\
\hline
HemodialysisRef3 & 214 & 54 & 154 & 422 & -35\%\\
\hline
LandingGearSystem\_3L & 46 & 19 & 18 & 83 & 17\%\\
\hline
OneWayTrafficLight & 72 & 9 & 56 & 137 & 261\%\\
\hline
PetriNet & 8 & 1 & 8 & 17 & 325\%\\
\hline
philosophers1 & 118800 & 1 & 118800 & 237601 & 1397553\%\\
\hline
Roulette & 5404 & 2703 & 2702 & 10809 & 77107\%\\
\hline
SluiceGateMotorCtl & 20 & 9 & 8 & 37 & 32\%\\
\hline
StereoacuityRaff5 & 30 & 7 & 16 & 53 & 2\%\\
\hline
\hline
AVG & 9688.69 & 217.08 & 9468.77 & 19374.54 & 25759\%\\
\bottomrule
\end{tabular}
\caption{Size of the flattened models}
\label{table:sizeFlattened}
\end{table}
The table also reports the total number of rules and their percentage change w.r.t. the original model ($\Delta$). We can observe that usually the flattened model has many more rules. The model that has the greatest increment in the number of rules is philosophers1; indeed, the model has several dynamic function arguments that, when flattened by \ar and \lr, produce several rules (see Table \ref{table:appliedFlatteners}).

However, there are also some models for which the number of rules is similar or also decreases; these models are already quite {\it flatten}: for example, the original model of HemodyalisisRef3 already contains almost only update, parallel, and conditional rules (see Table~\ref{table:benchSize}), and the application of the flattener has the effect of reducing the conditional rules by merging some of them through \nr (see Table~\ref{table:appliedFlatteners}).

We can interpret $\Delta$ as an index of the {\it conciseness} of the model: the higher $\Delta$ is, the more concise the original model is. Indeed, a very concise model (as philosophers1) specifies, by using few powerful rules, a complex behavior; when flattened, this results in a big number of rules.

\researchquestion{Which is the computational effort required by the flattener?}
To answer this question, we performed 100 executions of the flattener over all the models on a macOS machine, 3.1 GHz Intel Core i5, and 16GB. Table~\ref{table:appliedFlatteners} reports, for each model, the average time (in seconds) taken by the flattener, and the average time among models. We observe that for almost all models the execution time is less than 1.5 secs. We can notice that the model for which it takes longer (59.93 secs for philosophers1) is a very concise model that has been flattened a lot (see Tables~\ref{table:appliedFlatteners} and \ref{table:sizeFlattened}).

\section{Related work}\label{sec:related}

Flattening a model in order to simplify it is a rather common activity. The authors in~\cite{stateMachineFlattenSurvey} present flattening transformations for state machines equipped with hierarchy and parallelism, in order to transform models into executable code or inputs for model-based testing and verification techniques.

Model checkers use flatteners to simplify the notation of the models, as for example the NuSMV flattener~\cite{nusmvCAV2002}, that produces a synchronous flat model from a modular description of a model.

Model flattening has been exploited also in~\cite{flattenerB} to reduce model to code, in particular to generate efficient C code from B formal models in the domain of smart card applications.

In the context of the ASMs, Winter~\cite{Winterjucs} proposed some ad-hoc transformations for mapping ASMs to SMV models; differently, our flattener aims at producing a normal form of the model more widely usable for transformation to several other tools. A different ASM model refactoring approach appears in~\cite{refactorerASM} where a number of refactoring patterns are presented to restructure ASM models with the goal of improving their intelligibility, maintainability, abstraction, and conciseness. In a way, applying such patterns produces an effect which is opposite w.r.t. our flattener transformations: the latter ones may increase the model size, and may compromise model intelligibility; on the other side, they provide a normal form of the model in term of a very limited number of rule constructors, and the flattened model is not to be intended for readability and comprehension, but to facilitate tools integration.

\section{Conclusions}\label{sec:conclusions}

In this paper, we propose a \emph{flattener} tool, \asmetaF, integrated in the \asmeta framework that transforms an ASM in a \emph{flattened} model constituted only of update, parallel and conditional rules. The goal of the flattener is to support a pre-processing of ASM transformations towards tools having less expressive notational constructs, as, for example, those of verification tools. We claim that ASM flattening can improve the strengths of \asmeta as formal integrated development environment for ASMs.

In ASMs, functions are total by assigning the {\tt undef} value to undefined locations. In the translators developed in the past, handling the {\tt undef} has been challenging, and only some partial solutions (i.e., only for some domains) have been proposed. As future work, we plan to devise a special transformation able to produce an ASM in which the {\tt undef} can not occur and it is handled explicitly in the model.

As future step, we also plan to include in our flattener suitable transformations to handle \texttt{extend} and \texttt{seq} rules. This would allow reducing in normal form also advanced ASM classes that would become accessible to verification tools not yet able to support some classes of models.

\bibliographystyle{eptcs}
\bibliography{flattenerFIDE2018proc}
\end{document}